\documentclass[onecolumn,12pt,draftclsnofoot,peerreview,oneside]{IEEEtran}
\usepackage{mydefs}

\begin{document}

\title{A Pulse-Shape Binary Multiplex Modulation}

\author{Pavel Loskot, \IEEEmembership{Senior Member, IEEE}%
  \thanks{The author is with ZJU-UIUC Institute, Haining, China
    (e-mail: pavelloskot@intl.zju.edu.cn).}
  \thanks{This work was supported by a research grant from Zhejiang
    University.}}
\markboth{Pulse-Shape Binary Multiplex Modulation}{%
  Pulse-Shape Binary Multiplex Modulation}
\maketitle

\begin{abstract}
  The root raised-cosine pulse commonly used in linear digital
  modulations yields exactly two intersymbol interference components from the
  preceding and the subsequent data symbols, provided that the roll-off factor
  is $100\%$ and the modulation packing factor is set to $50\%$. This can be
  exploited to symmetrically multiplex two data streams of transmitted symbols.
  Hence, the proposed scheme is referred to as pulse-shape binary multiplex
  modulation. The demodulation of the two multiplexed data streams at the
  receiver can be aided by making the streams mutually orthogonal. It can be
  achieved by superposition modulation with symbol-by-symbol interference
  cancellation, proper design of transmission sequences interleaving pilot and
  data symbols in order to also enable channel estimation, and using orthogonal
  spreading sequences. The presented numerical results indicate that the
  proposed modulation scheme can outperform Nyquist signaling in terms of
  transmission reliability or the time required for transmitting the whole
  sequence of data symbols. For instance, differentially encoded modulation
  symbols can be transmitted twice as fast by the proposed modulation scheme
  with a 3 dB penalty in signal-to-noise ratio over additive white Gaussian
  noise channels.
\end{abstract}

\begin{IEEEkeywords}
  Intersymbol-interference; linear modulation; Nyquist signaling; partial
  response signaling; root raise cosine pulse; sequence multiplexing.
\end{IEEEkeywords}

\section{Introduction}

The spectrum scarcity necessities the use of spectrally efficient modulations.
The Nyquist signaling is a well established and robust technique for
constructing linear digital modulations which are employed in a vast majority
of today's communication systems. These modulation schemes are often combined
with channel encoding to improve the transmission reliability and even approach
the channel capacity. An alternative strategy is to assume modulations having a
controlled level of intersymbol interference (ISI), which can increase the rate
of information transmission as well as act as a form of information encoding
for improving the transmission reliability \cite{zhou2012}, albeit at the cost
of increased detection complexity at the receiver. Such so-called
faster-than-Nyquist (FTN) schemes are linear modulations that can be used over
band-limited channels \cite{anderson2013}.

The renewed interest in FTN signaling schemes goes back to the early 2000's
\cite{liveris2003}. However, a closely related idea of partial response linear
modulations with controlled ISI appeared much earlier \cite{proakis2008}. The
achievable spectral efficiency of coded and uncoded FTN schemes is evaluated in
\cite{anderson2013}, \cite{landau2017}, and \cite{modenini2012}. The
observation that up to $25\%$ increase in the transmission rate is possible
without deteriorating the error performance is known as the Mazo limit
\cite{liveris2003, anderson2013, fan2017, modenini2012}. The energy and
complexity costs of FTN signaling are reviewed in \cite{anderson2013}.

The FTN schemes can be implemented both in time and in frequency domains
\cite{anderson2013, yamada2015}. An orthogonal FTN scheme based on OFDM was
designed in \cite{bogale2017}. Alternatively, Nyquist signaling with dual root
raised-cosine (RRC) pulses akin to duobinary modulation has been investigated
in \cite{zhang2020}. This scheme was further refined for the RRC pulses with
zero roll-off in \cite{li2022}. The link between duobinary modulation and FTN
signaling has been pointed out in \cite{zhou2012}.

An important issue is how to efficiently perform the detection of transmitted
symbols at the receiver. Unlike the ISI due to multipath propagation, the ISI
created by FTN signaling also correlates samples of additive noise. The optimum
detection necessitates the use of whitening matched filter (WMF) prior to
symbol decisions. The ISI at the detector input can be equivalently represented
as an auxiliary channel \cite{landau2017, modenini2012}. The output signal of
such channel has a trellis-like structure, which can be optimally equalized by
the Viterbi, BCJR and other such algorithms with varying complexity
\cite{anderson2013, fan2017, modenini2012}. These decoding methods can approach
the performance of zero-ISI (Nyquist) modulations over additive white Gaussian
noise (AWGN) channels \cite{anderson2013}. The symbol-by-symbol detector for
FTN signals was devised in \cite{modenini2012} and \cite{bedeer2017}. The
detection of FTN signals with oversampling and one-bit quantization was
developed in \cite{landau2017}. A low complexity linear equalization for FTN
signaling was designed in \cite{bas2020}. The joint channel estimation and
decoding of FTN signals was studied in \cite{shi2018} and in \cite{wu2017}.

Nearly all investigations of FTN signaling schemes in the literature assume the
RRC modulation pulse. The RRC pulse is parameterized by a time period, $\Tp$,
and a roll-off factor, $\alpha$. Linear modulations combine the RRC pulses
weighted by data symbols, which are then transmitted once every symbol period,
$\Ts$. The packing factor defines the relationship between $\Tp$ and $\Ts$,
i.e., $\tau=1-\Ts/Tp$. The design and analysis of FTN signaling in the
literature usually assumes arbitrary values of $0\leq\alpha\leq 1$ and
$0\leq\tau< 1$. The search for good values of $\alpha$ and $\tau$ over an
entire $\alpha-\tau$ plane to allow symbol-by-symbol decisions was carried out
in \cite{bedeer2017}. However and importantly, the case of $100\%$ bandwidth
roll-off is rarely explicitly considered in the literature \cite{modenini2012}.
The authors in \cite{landau2017} noticed that, for $\alpha=1$ and an arbitrary
value of $\tau$, the ISI is approximately limited to the two previous and the
two subsequent symbol samples.

In this paper, we show that the RRC pulse with $100\%$ roll-off and $50\%$
packing has a well-defined ISI, which is exactly and symmetrically constrained
to one previous and one subsequent symbol. Such a unique property of the RRC
pulse appears to remain unnoticed in the literature. Interestingly, reference
\cite{zhou2012} states that ISI with only two components can be obtained with
$100\%$ roll-off and $50\%$ packing assuming prolate spheroidal wave pulses,
but not RRC pulses. Although such a modulation scheme can be assumed to be a
special case of FTN signaling, it is argued that RRC pulses having $100\%$
roll-off and $50\%$ packing offers symmetric multiplexing of the two
transmitted data streams. For this reason, such a partial response signaling is
referred to in this paper as a pulse-shape binary multiplexing (PSBM)
modulation. The main task then is how to separate the two multiplexed data
streams at the receiver with acceptable reliability and complexity. As with
other partial response signalings, the modulation constellation and the
dependency between transmitted symbols must be carefully selected in order to
trade-off the performance and the decoding complexity. We design several
transmission sequences interleaving pilot and data symbols, discuss
superposition modulation with symbol-by-symbol sequential interference
cancellation (SIC), and also consider orthogonal spreading sequences to aid
separation of the data streams at the receiver. In addition, the performance of
multiplexed differentially encoded phase-shift keying (PSK) modulation symbols
is evaluated numerically. The numerical results identify several cases when the
proposed PSBM modulation outperforms the Nyquist signaling in terms of either
transmission reliability or the time required to transmit a given number of
data symbols.

The rest of this paper is organized as follows. Linear modulation schemes that
are related to the proposed pulse-shape multiplexing signaling are outlined in
Section II. System model and the received signal structure are described in
Section III. The proposed pulse-shape multiplexing modulation is defined in
Section IV including the design of transmitted symbol sequences. Numerical
results are presented in Section V. Section VI concludes the paper.

We adopt the following notations: $\E{\cdot}$ is expectation, $\conv$ is
convolution, $|\cdot|$ is absolute value, $(\cdot)^\ast$ is complex conjugate,
$\real{\cdot}$ and $\imag{\cdot}$, respectively, denote the real and imaginary
part of a complex number, $\Card\{\cdot\}$ is cardinality of a set, $(\cdot)^T$
is matrix transpose, $(\cdot)^{-1}$ is matrix inverse, and $\norm{\cdot}$ is
the Euclidean norm of a matrix or vector.

\section{Related Linear Modulation Schemes}

A linearly modulation signal is constructed as,
\begin{equation}\label{eq:5}
  x(t) = \sum_k s_k \, p(t-k \Ts)
\end{equation}
where $s_k$ are $M$-ary modulation symbols transmitted every symbol period,
$\Ts$, and $p(t)$ denotes a deterministic pulse-shape, which is also known at
the receiver. The stationary sequence of transmitted symbols, $s_k$, has
zero-mean, and the variance, $\E{|s_k|^2}=\Es$. The symbols are usually
obtained as output of a finite-state modulator, i.e.,
\begin{equation}\label{eq:6}
  s_k = s(q_k,c_k)
\end{equation}
where the states, $q_k$, represent modulation memory, and the data symbols,
$c_k$, each carry, $\log_2 M$, bits of input information. In this paper, $p(t)$
is assumed to be the unit-energy RRC pulse, \cite{proakis2008}
\begin{equation}\label{eq:7}
  p(t) = \frac{\rrc_\alpha(t/\Ts)}{\sqrt{\Ts}} 
\end{equation}
where 
\begin{equation}\label{eq:7a}
  \rrc_\alpha(t) = \frac{1}{1-16\alpha^2 t^2}
  \left( \frac{\sin\left((1-\alpha)\pi t\right)}{\pi t}
    + \frac{4\alpha\cos\left((1+\alpha)\pi t \right)}{\pi}\right).
\end{equation}
The roll-off factor, $0\leq \alpha \leq 1$, however, it is possible to also
consider pulse shapes having a roll-off greater than $100\%$.

Since the sequence of symbols, $s_k$, is stationary, the auto-correlation,
$R_s(i-j) = \E{s_i s_j^\ast}$. The corresponding power-spectrum density (PSD)
of signal \eref{eq:5} is computed as, \cite{proakis2008}
\begin{equation}\label{eq:26}
  S_x(f) = \frac{1}{\Ts} |P(f)|^2 \ \sum_k R_s(k) \eee^{\jj 2\pi f k \Ts}
\end{equation}
where $P(f)$ denotes the Fourier transform of $p(t)$.

Correlative coding assumes the discrete modulator \eref{eq:6} to be a finite
impulse response (FIR) filter, i.e.,
\begin{equation}\label{eq:60}
  s_k = \sum_{i=0}^{K-1} v_i c_{k-i}.
\end{equation}
The filter weights, $v_i$, are normalized, so that, $\sum_i |v_i|^2=1$. More
importantly, with a change in indices, modulated signal \eref{eq:5} with
symbols \eref{eq:60} can be rewritten as,
\begin{equation}
  \begin{split}
    x(t) =& \sum_k \sum_{i=0}^{K-1} v_i\, c_{k-i} p(t - k\Ts) \\ =&
    \sum_k c_k \sum_{i=0}^{K-1} v_i\, p(t-(k+i)\Ts) = \sum_k c_k\, \tp(t-k\Ts)
  \end{split}
\end{equation}
where the compound pulse, $\tp(t)= \sum_{i=0}^{K-1} v_i \, p(t-i\Ts)$.

Duobinary modulation is a special case of correlative coding, such that the FIR
filter has only two non-zero weights, $v_o=v_1=1/\sqrt{2}$, the modulation
symbols are binary, i.e., $c_k \in \{-\sqrt{\Es},+\sqrt{\Es}\}$, and the RRC
pulse has the smallest possible roll-off, $\alpha=0$. Modified duobinary
modulation assumes instead the weights, $v_0=1/\sqrt{2}$, $v_1=0$, and
$v_2=-1/\sqrt{2}$.

The following modulations assume the RRC pulse-shape with an arbitrary roll-off
value. Differential PSK constructs the transmitted symbols as,
\begin{equation}\label{eq:13}
  s_k = c_k\,s_{k-1}
\end{equation}
where the data symbols, $c_k\in\{ \sqrt{\Es} \eee^{\jj 2\pi(i-1)/M} \}$,
$i=1,2,\ldots,M$. Generalized shift-keying extends the modulation alphabet of
amplitude or phase shift-keying modulations with a zero symbol
\cite{loskot2012}. Offset-quadrature ($M=4$) PSK delays the imaginary part of
the modulated signal by half a symbol period, i.e.,
\begin{equation}\label{eq:39}
  x(t) = \sum_k \real{c_k} p(t-k\Ts) + \jj\, \imag{c_k} p(t-k\Ts-\Ts/2).
\end{equation}

Finally, FTN signaling is a linear modulation described by eq. \eref{eq:5}.
More importantly, the RRC pulse-shape in \eref{eq:7} can now be scaled by,
$\Tp= \Ts/(1-\tau)$, instead of $\Ts$, where $0\leq \tau < 1$ is so-called the
packing factor, i.e.,
\begin{equation}\label{eq:29}
  x(t)= \sum_{k} s_k p(t-k(1-\tau)\Tp)=\sum_{k} s_k p(t-k \Ts)
\end{equation}
so that $\Tp$ is a design parameter of the pulse, $p(t)$, whereas,
$\Ts=(1-\tau)\Tp$, denotes the symbol period. Thus, $\tau=0$ packing
corresponds to a conventional Nyquist signaling, whereas $\tau=1$ packing would
completely overlap the transmitted symbols. More importantly, the PSD of
\eref{eq:29} is still given by eq. \eref{eq:26}, and it is otherwise completely
independent of the packing factor, $\tau$.

\section{Received Signal}

The standard wireless channel model with $L$ propagation paths is an FIR filter
with the impulse response,
\begin{equation}\label{eq:12}
  \tth(t) = \sum_{l=1}^L h_l(t) \delta(t - \tau_l).
\end{equation}
The signal delays, $\tau_l$, are assumed to be constant. The path attenuations,
$h_l(t)$, are zero-mean circularly symmetric Gaussian processes. These
processes are stationary, and generally mutually correlated. They have a
defined auto-correlation, $R_h(\Delta t)$, which determines the coherence
bandwidth.

For narrow-band signals, the number of paths, $L$, is small. For $L=1$, the
channel model \eref{eq:12} becomes frequency non-selective. In low-mobility
scenarios, the channel attenuations, $h_l(t)$, are often assumed to be constant
over blocks of transmitted symbols, and independent between the successive
blocks, which is often referred to as a block fading model.

The received signal corresponding to multi-path propagation model \eref{eq:12}
is written as,
\begin{equation}
  \begin{split}
    y(t) &= \tth(t) \conv x(t) + w(t) \\
    &= \sum_{l=1}^L h_l(t) x(t-\tau_l) + w(t)
  \end{split}
\end{equation}
where $w(t)$ is a zero-mean stationary circularly symmetric AWGN with the
variance, $\sgw^2=\E{|w(t)|^2}$.

The received signal is filtered through a filter matched to the transmitted
pulse, $p(t)$, and synchronously sampled at a rate, $1/\Ts$. In particular,
assuming RRC pulses, the matched filter, $p^\ast(-t)=p(t)$, and provided that
the channel attenuations are constant over blocks of transmitted symbols, the
received samples are modeled as,
\begin{equation}\label{eq:17}
  \begin{split}
    r_n=& y(t)\conv p^\ast(-t)\Big|_{t=n\Ts+\tau_0} \\ =& \sum_{l=1}^L h_l\,
    x(t-\tau_l)\conv p(t)\Big|_{t=n\Ts+\tau_0} +
    w(t)\conv p(t)\Big|_{t=n\Ts+\tau_0}\\
    =& \sum_{l=1}^L h_l \sum_k s_k \int_{-\infty}^\infty
    p(\zeta+(n-k)\Ts-\tau_l) p(\zeta-\tau_0) \df \zeta \\ &+
    \int_{-\infty}^\infty w(\zeta+n\Ts) p(\zeta-\tau_0) \df \zeta \\
    =& \sum_k s_k \sum_{l=1}^L h_l \, p_{n-k,l} + w_n = \sum_k s_k
    \tp_{n-k} + w_n \\ =& s_k\tp_0 + \underbrace{\sum_{k \atop n\neq k} s_k
      \tp_{n-k}}_{\xISI} + w_n.
\end{split}
\end{equation}
The timing offset, $\tau_0$, at the receiver can be optimized to minimize the
ISI term (in some sense) in \eref{eq:17} defined as,
\begin{equation}\label{eq:18}
  \tp_{n-k} = \sum_{l=1}^L h_l \int_{-\infty}^\infty
  p(\zeta+(n-k)\Ts-\tau_l) p(\zeta-\tau_0) \df \zeta,\ n\neq k.
\end{equation}
Thus, the ISI arises when the orthogonality between the transmitter and the
receiver pulses is violated, for example, due to multi-path propagation,
time-synchronization errors between transmitter and receiver, and also due to
symbol-period compression in FTN signaling schemes \cite{proakis2008}.

An interesting question is how much ISI is produced for different combinations
of parameters $\alpha$ and $\tau$ in FTN signaling schemes using RRC pulses.
Hence, define the function,
\begin{equation}
  \ISI{\mu} = \Card \{ |\tp_k|>\mu,\ k\neq 0\}
\end{equation}
to be the number of ISI components that are greater than a given threshold,
$\mu$. Note that, $\ISI{\mu}\in\{0,2,4,\ldots\}$, due to even symmetry of the
RRC pulses. Assuming different thresholds, $\mu$, the roll-off,
$0\leq \alpha\leq 2$, and the RRC pulses truncated to $(-4\Ts,+4\Ts)$, the
values $\ISI{\mu}=0$ (red points) and $\ISI{\mu}=2$ (blue points) in the
$\alpha-\tau$ plane are shown in \fref{pict01}. The empty (white) spaces in
\fref{pict01} indicate the values, $\ISI{\mu}>2$. It can be observed that by
decreasing the threshold, $\mu$, several cases of interest for designing FTN
signaling schemes start to emerge. In particular, exactly two ISI components
can be obtained for these parameters: $\alpha=1.0$ and $\tau=0.5$,
$\alpha=1.07$ and $\tau\in(0.70,0.71)$, and $\alpha\in(1.65,1.85)$ and
$\tau\in(0.47,0.50)$.

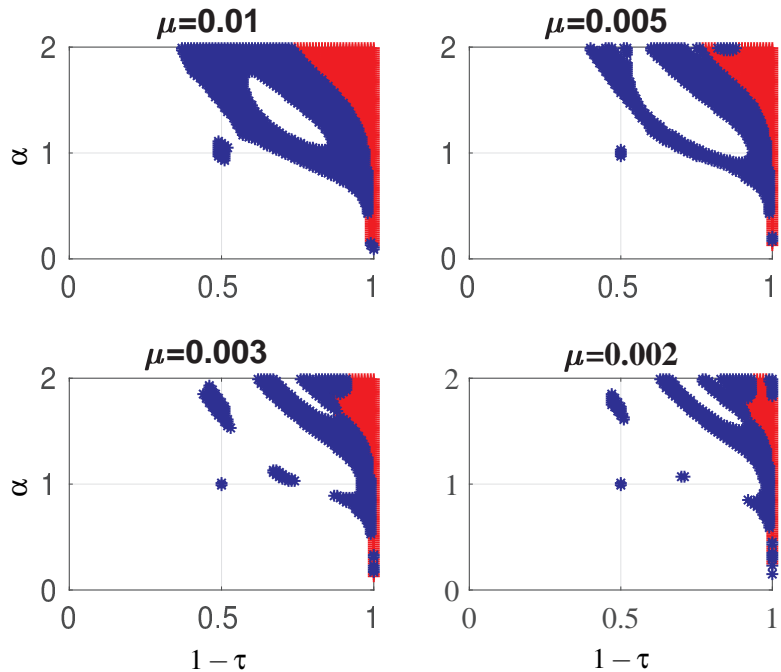
\insfig{0}{scale=1.0}{pict01}{The $\ISI{\mu}=0$ components (red points) and
  $\ISI{\mu}=2$ components (blue points) for four different thresholds, $\mu$.}

\section{Pulse-Shape Binary Multiplex Modulation}

As indicated in \fref{pict01}, the RRC pulse with $100\%$ roll-off and $50\%$
packing has well-defined and finite ISI components. In particular, the RRC
pulse \eref{eq:7a} for $\alpha=1$ becomes,
\begin{equation}
  \rrc_1(t) = \frac{4 \cos(2\pi t)}{\pi(1-16 t^2)}.
\end{equation}
This pulse has the following ISI components in an AWGN channel without
multi-path. Such a fundamental property appears to remain unnoticed in the
literature.

\begin{lemma}\label{lm:1}
  Let $n$ be a non-negative integer. The ISI integral involving the RRC pulse,
  $\rrc_1(t)$, with $100\%$ roll-off has the exact solution,
  \begin{equation}\label{eq:33}
    \begin{split}
      \int_{-\infty}^\infty & \rrc_1(t) \times \rrc_1(t-n/4)\df t \\ =& \left\{
        \begin{array}{cc} 8/(3\pi) & n=1 \\
          \frac{8}{\pi(n-2)n(n+2)} & n-\mathrm{odd},\ n>1 \\
          1 & n=0 \\ 1/2 & n=2 \\ 0 & n-\mathrm{even},\ n>2 .
        \end{array} \right.
  \end{split}
\end{equation}
\end{lemma}
\lmref{lm:1} can be proved by solving the integral for the first few values of
$n$ (for example, using \textsf{Mathematica} software), and then using
induction.

However, the result \eref{eq:33} is exact only when the integration is
performed over an infinite interval. In practice, the pulse shapes must be
truncated to a finite interval. The numerically computed values of integral
\eref{eq:33} when the interval of integration is truncated to $(-d,+d)$ are
shown in \fref{pict02}. It can be observed that the RRC pulse shape,
$\rrc_1(t)$, should not be truncated to the intervals shorter than, $(-4,+4)$,
in order to achieve the RRC property given in \lmref{lm:1} with at least
$99.9\%$ accuracy.

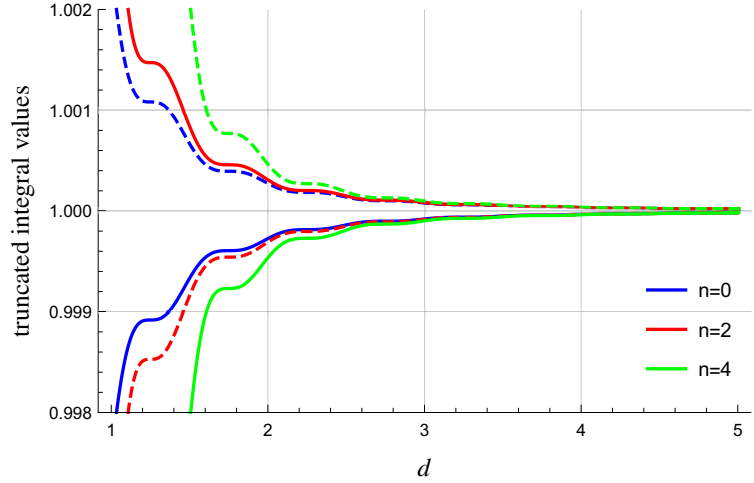
\insfig{0}{scale=0.9}{pict02}{Numerically computed integral \eref{eq:33}
  truncated to interval, $(-d,+d)$, as a function of $d$ (solid lines). The
  exact values for an infinite interval are shifted to be all equal to unity in
  order to enable comparison. The dashed lines are mirrored solid lines about
  the unit value.}

\begin{definition}\label{df:1}
  The modulated signal of pulse-shape binary multiplex modulation is written
  as,
  \begin{equation}\label{eq:21}
    x(t) = \sum_k s_k \frac{\rrc_1\left(\frac{t-k\Ts}{2\Ts}\right)
    }{\sqrt{2\Ts}}.
  \end{equation}
\end{definition}
  
The synchronously sampled matched filter output of modulated signal
\eref{eq:21} received in AWGN, $w(t)$, is,
\begin{equation}\label{eq:23}
  \begin{split}
    r_n =& (x(t)+w(t)) \conv \frac{\rrc_1\left(\frac{t}{2\Ts}\right)}
    {\sqrt{2\Ts}}\Big|_{t=n\Ts} \\ =& \sum_k s_k \int_{-\infty}^\infty 
    \rrc_1(\zeta + (n-k)/2) \ \rrc_1(\zeta) \df\zeta + w_n \\
    =& \left( \frac{1}{2} s_{n-1} + s_n + \frac{1}{2} s_{n+1} \right)
    + w_n.
  \end{split}
\end{equation}
  
The noise samples, $w_n$, in \eref{eq:23} are zero-mean, have the variance,
$\E{|w_n|^2}= \E{|w(t)|^2}=\sigma_w^2$, and their stationary auto-correlation
is,
\begin{equation}\label{eq:37}
  R_w(n-m) = \E{w_nw_m^\ast} = \left\{ \begin{array}{cc} \sigma_w^2 & n=m \\
      \sigma_w^2/2 & |n-m|= 1 \\ 0 & |n-m|>1. \end{array} \right.
\end{equation}
Such noise samples can be equivalently modeled by a simple FIR filter,
\begin{equation}
  w_n = \frac{u_n+u_{n-1}}{\sqrt{2}}
\end{equation}
where $u_n$ are the samples of a zero-mean, circularly symmetric Gaussian
process having the variance, $\E{|u_n|^2}=\sigma_w^2$. In addition, it is
straightforward to show that the variance of the sum of $N$ noise samples
having the correlations \eref{eq:37} is,
\begin{equation}
  \var{\sum_{n=1}^N w_n} = (2N-1)\sigma_w^2
\end{equation}
which is greater than the variance, $N\sigma_w^2$, of the sum of $N$
uncorrelated samples.

The modulated signal \eref{eq:21} in \dfref{df:1} can be visualized as shown in
\fref{pict03}. In particular, the transmitted data symbols can be viewed as
consisting of two multiplexed streams of data symbols, $a_k$, and, $b_k$, which
are each transmitted with a period $2\Ts$, but mutually shifted by $\Ts$. The
corresponding received symbol samples after the matched filtering are,
\begin{equation}\label{eq:34}
  r_n = \left\{ \begin{array}{cc} a_n + \frac{b_{n-1}+b_n}{2} + w_n &
      n-\mbox{odd} \\ b_n + \frac{a_n + a_{n+1}}{2} + w_n &
      n-\mbox{even}. \end{array} \right.
\end{equation}

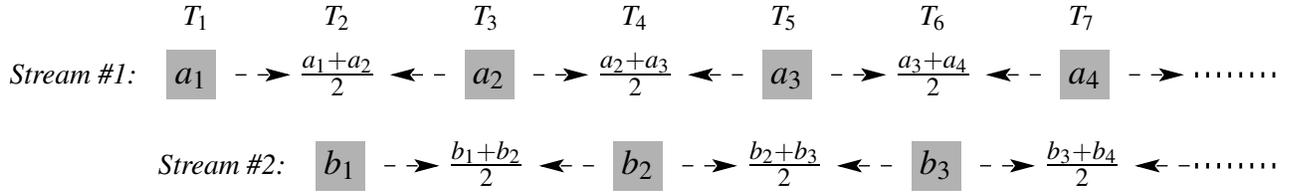
\insdfig{0}{scale=1.1}{pict03}{A visualization of pulse-shape multiplex
  modulated signal.}

Using \eref{eq:26}, the PSD of modulated signal \eref{eq:21} is computed as,
\begin{equation}
  S_x(f) = 2 |\RRC_1(2\Ts f)|^2 \, \sum_k \E{s_0s_k^\ast}
  \eee^{\jj 2\pi f k \Ts}
\end{equation}
where the Fourier transform of the pulse, $\rrc_1(t)$, is,
\begin{equation}
  \RRC_1(f) = \left\{ \begin{array}{cc} \cos(\pi f/2) &
      |f| \leq 1 \\ 0 & \mbox{otherwise}. \end{array} \right.
\end{equation}

\subsection{Transmitted Sequence Design}

The optimum detection of transmitted symbols in the presence of ISI must
consider complete sequences of received samples. However, in the absence of
multi-path, the received samples have structure \eref{eq:23}, and the
transmitted sequences can be designed, so that the complexity of detection at
the receiver can be reduced.

The key strategy for reducing the detection complexity is to exploit
orthogonality among sub-sequences of transmitted symbols. Offset-quadrature PSK
modulation \eref{eq:39} alternates one-dimensional modulation symbols along the
in-phase and quadrature components, which allows the optimum symbol-by-symbol
decisions.

Multiplexing two data streams as described by \eref{eq:34} can exploit the
design principles of superposition modulation and multiuser detection. In such
a case, symbol-by-symbol decisions can be performed by SIC. Specifically,
provided that symbols, $b_n$, can be reliably detected, even if the symbols,
$(a_n+a_{n+1})/2$, are not yet known, then the symbol, $a_n$, can be reliably
detected after canceling the ISI term, $(b_{n-1}+b_n)/2$.

In the sequel, three other sequence design strategies are discussed in more
detail. The first strategy combines pilot and data symbols to aid the data
detection and channel estimation. The second strategy employs orthogonal
spreading codes in order to separate the two multiplexed data sequences. The
third strategy adopts the differential encoding of transmitted symbols.

\subsection{Sequences with Interleaved Pilot Symbols}

In general, pilot symbols for channel estimation can be interleaved with data
symbols or superimposed onto data symbols \cite{jagannatham2006}. Here, the
more common former approach is adopted. Thus, consider a transmitted sequence
consisting of alternating groups of $\Ld$ data symbols and $\Lp\ll \Ld$ pilot
symbols, which are separated by a single zero-symbol as shown in \fref{pict04}.

For instance, the following sub-sequences with reduced or no ISI can be
considered with pilot symbol, $\pp$, and arbitrary data symbols, $\dd_1$, and,
$\dd_2$: $(0,\pp,0)$, $(\dd_1,\pp,-\dd_1)$, $(\dd_1,\pp,-\dd_1,-\pp,\dd_1)$,
and $(\dd_1,\pp,-\dd_1,-\pp,\dd_2,\pp,-\dd_2)$. These sub-sequences enable
ISI-free data detection and channel estimation, as can be deduced from eq.
\eref{eq:34} and \fref{pict03}. Recall also that the noise samples, $w_n$, and,
$w_{n\pm 2}$, are uncorrelated, i.e., independent.

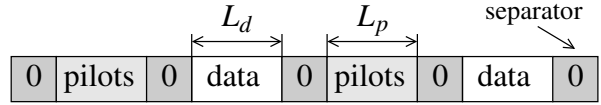
\insfig{0}{scale=1.0}{pict04}{The transmitted sequence with interleaved
  sub-sequences of data and pilot symbols and a single zero-symbol separator.}

In order to illustrate the ISI-free channel estimation, consider the sequence,
$(\dd_1,\pp,\pp,-\dd_1)$. The received samples corresponding to the two pilot
symbols in the middle are,
\begin{equation}
  \begin{split}
    r_n =& h \frac{3}{2}\pp+h \frac{1}{2} \dd_1+w_n \\
    r_{n+1} =& h \frac{3}{2}\pp - h \frac{1}{2} \dd_1+w_{n+1}
  \end{split}
\end{equation}
where $h$ denotes the complex-valued channel attenuation (i.e., frequency
non-selective slow fading). The samples, $r_n$, and, $r_{n+1}$, can be simply
combined as,
\begin{equation}
  r_n+r_{n+1}= 3 h \pp +w_n+w_{n+1}
\end{equation}
where the total variance of the additive noise samples is equal to
$3\sigma_w^2$ due to correlations \eref{eq:37}.

More generally, the transmitted sequence,
\begin{equation}
  (-\pp,\dd_1,\pp,\dd_2,-\pp,\dd_3,\pp,\dd_4,-\pp,\dd_5,\ldots,\dd_N,\pm \pp)
\end{equation}
where the last pilot symbol is $\pp$, if $N$ is odd, and $-\pp$, if $N$ is
even, allows the ISI-free symbol-by-symbol decisions of all data symbols.
Moreover, assuming again a slow fading channel, the received samples
corresponding to the pilot symbols can be summed up to obtain,
\begin{equation*}
  \sum_{n=1}^N (-1)^n \, r_{2n-1}= N\,h\,\pp+\sqrt{N} w
\end{equation*}
where the noise sample, $w$, has the variance, $\sigma_w^2$, so the
signal-to-noise ratio (SNR) for estimating the channel coefficient, $h$, has
been improved $N$-times. Note also that once the channel has been estimated,
the pilot symbols can be subtracted from the received samples in order to aid
decisions of the remaining data symbols.

Finally, consider the case of a symbol repetition diversity. The transmitted
sequence, $(\dd,0,\dd,0,\ldots,0,\dd)$, of a data symbol, $\dd$, repeated
$(N\geq 2)$-times corresponds to the canonical Nyquist signaling. The
pulse-shape multiplex modulation instead transmits the sequence,
$(\dd,\dd,\ldots,\dd)$, of $N_1$-times repeated data symbol, $\dd$. For the
same sequence length, $N_1=2N-1$. Assuming slowly fading channel, the detector
combines the received samples for the two modulation schemes, respectively, as,
\begin{equation}
  \begin{split}
    r =& N\, h\, d + \sqrt{N} w \\
    r =& (2\cdot3/2+2(N_1-2))\,h\, d/\sqrt{2}+\sqrt{2N_1-1} w
  \end{split}
\end{equation}
where the scaling by $\sqrt{2}$ was introduced for the second modulation in
order to account for the larger number of symbols in its transmitted sequence.
The resulting SNR of these two schemes is proportional to, $\gamma\propto N$,
and, $\gamma\propto 2N-3/2$, respectively. Consequently, for symbol repetition
diversity, the SNR gain of the pulse-shape binary multiplexing is
asymptotically $3$ dB larger than for the Nyquist signaling.

\subsection{Sequences with Orthogonal Spreading}

Another strategy for transmitting interleaved, but orthogonal symbols in
modulated signal \eref{eq:21} is to use orthogonal spreading codes. In
particular, assume transmitted symbols,
\begin{equation}
  a_n = \dd_1 c_n^{(1)},\qquad b_n = \dd_2 c_n^{(2)} 
\end{equation}
where $\dd_1$ and $\dd_2$ are two data symbols, and, $c_n^{(1)}$ and
$c_n^{(2)}$, $n=1,2,\ldots,N$, are generally complex-valued, orthogonal
spreading sequences, so that,
\begin{equation}
  \sum_{n=1}^N c_n^{(i)} c_n^{\ast(j)} = \left\{ \begin{array}{cc} N, & i=j \\
      0, & i\neq j. \end{array} \right.
\end{equation}
Then, the sequences of received samples \eref{eq:34} are linearly combined as,
\begin{equation}
  \begin{split}
    \sum_{n=1}^N r_{2n-1} c_n^{\ast(1)} =& \dd_1 + \dd_2 \sum_{n=1}^N
    \frac{c_n^{(2)}+c_{n+1}^{(2)}}{2}\, c_n^{\ast(1)} + \sum_{n=1}^N w_{2n-1}
    c_n^{\ast(1)} \\ =& \dd_1 + \tw_1 \\
    \sum_{n=1}^N r_{2n} c_n^{\ast(2)} =& \dd_2 + \dd_1 \sum_{n=1}^N
    \frac{c_n^{(1)}+c_{n+1}^{(1)}}{2}\, c_n^{\ast(2)} + \sum_{n=1}^N w_{2n}
    c_n^{\ast(2)} \\ =& \dd_2 + \tw_2
  \end{split}
\end{equation}
provided that the spreading sequences, $c_n^{(1)}$, and, $c_n^{(2)}$, are
exactly orthogonal. In such a case, the SNR improvement for transmitting two
data symbols with orthogonal spreading sequences using the pulse-shape binary
multiplex modulation \eref{eq:21} is proportional to,
\begin{equation}
  \gamma \propto \frac{N^2}{2N-1}.
\end{equation}

For instance, if the spreading symbols, $c_n$, are generated independently at
random and with an equal probability from the set, $\{-1,+1\}$, the probability
that two such sequences are orthogonal is,
\begin{equation}\label{eq:43a}
  \begin{split}
    \Prob{\sum_{n=1}^N c_n^{(1)} c_n^{\ast(2)}=0} =& \binom{N}{N/2}
    \left(\frac{1}{2}\right)^{N/2} \left(\frac{1}{2}\right)^{N-N/2} \\ =&
    \binom{N}{N/2} 2^{-N}.
  \end{split}
\end{equation}
Since the probability \eref{eq:43a} of exact orthogonality asymptotically goes
to zero with large $N$, consider instead the probability,
\begin{equation}\label{eq:43b}
  \begin{split}
    \Prob{-\lceil \kappa\,N/2 \rfloor \leq \sum_{n=1}^N c_n^{(1)} c_n^{\ast(2)}
      \leq \lceil \kappa\,N/2 \rfloor} \\= \sum_{n=-\lceil \kappa\,N/2
      \rfloor}^{\lceil \kappa\,N/2 \rfloor} \binom{N}{n} \left( \frac{1}{2}
    \right)^N
  \end{split}
\end{equation}
for some small $\kappa\geq 0$. The probabilities \eref{eq:43b} as a function of
$N$ for two different values of factor, $\kappa$, are shown in \fref{pict05}.
These probabilities are indicative of how many random spreading sequences need
to be generated in order to select the required number of such sequences having
an acceptable level of mutual orthogonality.

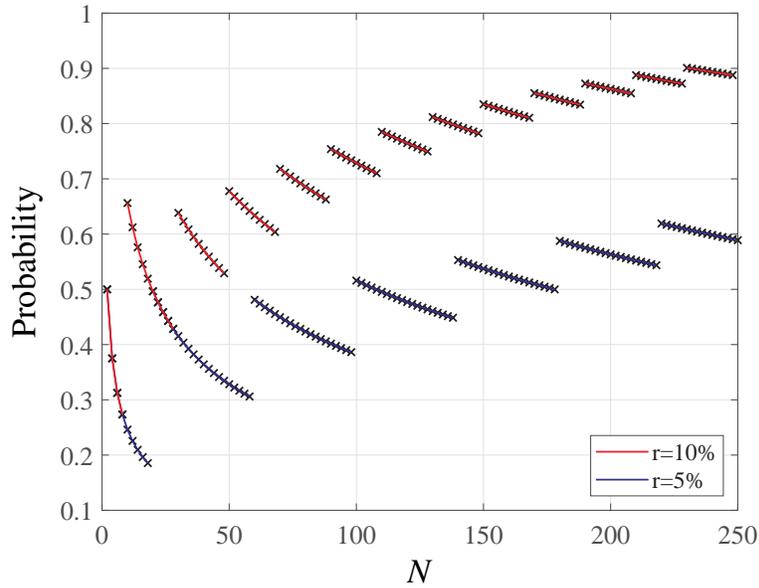
\insfig{0}{scale=1.2}{pict05}{The probability \eref{eq:43b} vs. the spreading
  sequence length, $N$, assuming $\kappa=5\%$ and $\kappa=10\%$, respectively.}

\subsection{Sequences with Differential Encoding}

Differential PSK is a popular modulation scheme for fast fading channels, since
it alleviates the need for recovering the absolute phase reference.
\fref{pict06} shows differentially encoded $M$-ary PSK symbols \eref{eq:13}
transmitted via pulse-shape binary multiplex modulation. In particular, the
$n$-th transmitted symbol is,
\begin{equation}
  \begin{split}
    s_n =& \left(\prod_{k=0}^{n-2} c_k\right) \frac{1+c_{n-1}c_n+2c_{n-1}}{2}
    \\ =& \frac{1}{2} \left(\prod_{k=0}^{n-2} c_k\right) +
    \frac{1}{2} \left(\prod_{k=0}^{n-1} c_k\right) c_n +
    \left(\prod_{k=0}^{n-1} c_k\right).
  \end{split}
\end{equation}
Consequently, the differential decoding can be performed as,
\begin{equation}
  \begin{split}
    c_n =& \left( 2 s_n - \left(\prod_{k=0}^{n-2}
        c_k\right) - 2 \left(\prod_{k=0}^{n-1} c_k\right) \right)
    \left(\prod_{k=0}^{n-1} c_k\right)^\ast \\
    =& 2 s_n \left(\prod_{k=0}^{n-1} c_k^\ast \right) - c_{n-1}^\ast -2.
  \end{split}
\end{equation}
The performance of this modulation scheme is evaluated in the next section.

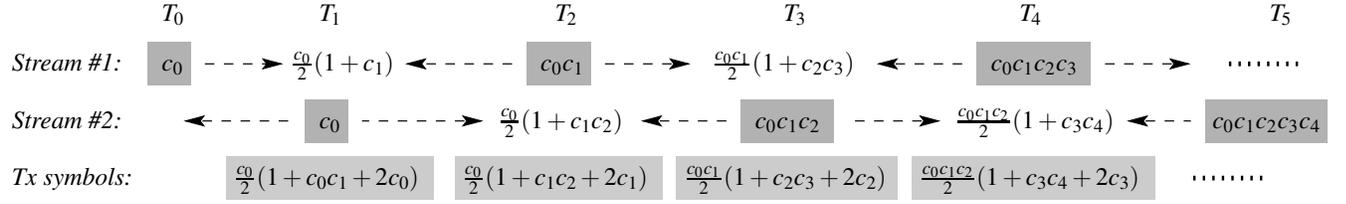
\insdfig{0}{scale=0.95}{pict06}{Differentially encoded $M$-ary PSK symbols
  transmitted via pulse-shape binary multiplex modulation.}

\section{Numerical Examples}

It is convenient to use a vector notation to generate samples of pulse-shape
binary multiplex modulation \eref{eq:21} received over a frequency
non-selective fading channel. The vector, $\vr$, of $N$ received samples
corresponding to the vector, $\vs$, of $N$ transmitted symbols can be obtained
as,  
\begin{equation*}
  \vr= \vs\,\vA\,\diag{\vh}+\vw\,\vA_0^T
\end{equation*}
where $\vh$ is a vector of fading channel coefficients, $\vw$ are samples of
AWGN, and the $(N\times N)$ ISI matrix,
\begin{equation*}
  \vA = \left[ \begin{matrix} 1 & 1/2 & & \\ 1/2 & 1 & 1/2 & \\
      & \ddots & \ddots & \\ & & 1/2 & 1 \end{matrix} \right] = \vA_0\,\vA_0^T.
\end{equation*}

The optimum detection requires that the additive noise is first whitened as,
\cite{proakis2008}
\begin{equation*}
  \vr\,\vA_0^{-T}= \vs\,\vA\,\diag{\vh}\,\vA_0^{-T}+\vw.
\end{equation*}
Then the maximum likelihood (ML) detection of sequence $\vs$ is,
\begin{equation}\label{eq:50}
  \hat{\vs}= \arg\min\limits_{\vs}
  \norm{\vr\,\vA_0^{-T}-\vs\,\vA\,\diag{\hat{\vh}}\,\vA_0^{-T}}^2
\end{equation}
where $\hat{\vh}$ is the estimate of $\vh$ representing channel state
information (CSI).

An uncoded binary phase shift keying (BPSK) modulation and Rayleigh-distributed
fading amplitudes, $\vh$, are assumed for simplicity. The transmitted sequence
interleaves pilot symbols and data symbols as shown in \fref{pict04}. The pilot
symbols are used to estimate the channel coefficients, $\vh$, by linear minimum
mean-square error (LMMSE) algorithm. The spectral efficiency of pulse-shape
binary multiplexing is, $2$, which is always larger than the spectral
efficiency of the Nyquist signaling being equal to, $2/(1+\alpha)$.

The BER curves, $\Pe$, for short data sequences of $\Ld=4$ and $\Ld=8$ binary
symbols, respectively, separated by a single zero-symbol are shown in
\fref{pict07} and \fref{pict08}. The SNR is defined as,
$\gamma_b= 1/(2\sgw^2)$. Both cases of perfect and estimated CSI are
considered. The Nyquist signaling (no ISI) with symbol-by-symbol decisions is
assumed as a reference. The ML data detector \eref{eq:50} is used for
pulse-shape multiplexing signaling.

It can be observed that the performance penalty due to channel estimation is
much larger for pulse-shape multiplexing than for the Nyquist signaling, which
is to be expected. The WMF improves the performance by several dB's for both
signaling schemes. More importantly, the performance of pulse-shape
multiplexing improves with the data block length by exploiting the time
diversity over a fading channel, so it can significantly outperform the Nyquist
signaling at medium to large SNR values. It is likely that by employing more
sophisticated channel estimation and equalization techniques, the performance
of pulse-shape multiplexing can be further improved. In order to demonstrate
the effect of time diversity, \fref{pict09} shows that, over an AWGN channel,
the performance of pulse-shape multiplexing is worse than that of Nyquist
signaling, even though some performance loss can be recovered by WMF.

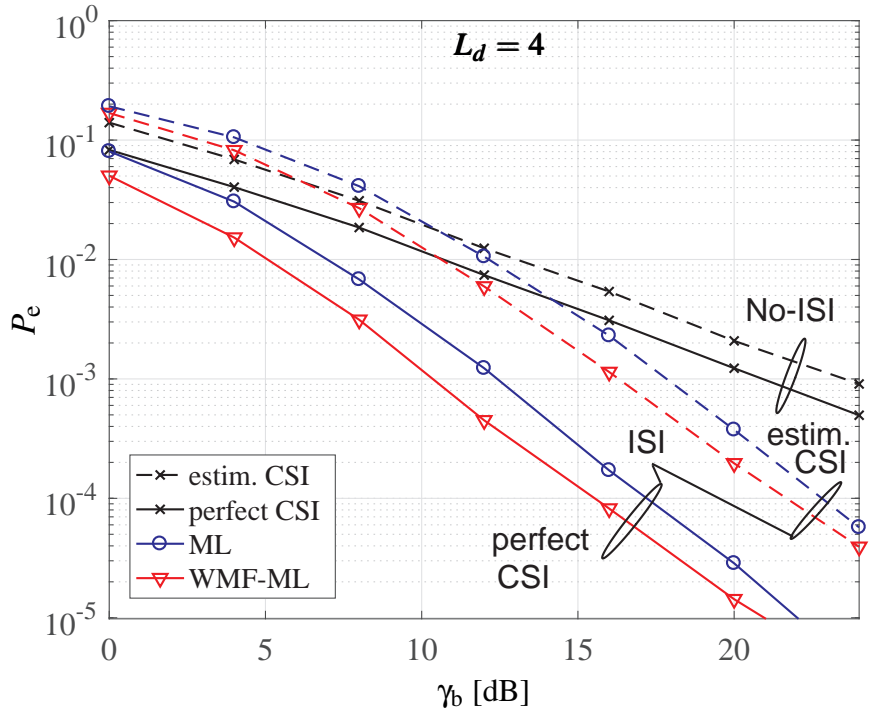
\insfig{0}{scale=1.0}{pict07}{The BER of BPSK vs. SNR over Rayleigh fading
  channel for sequences of $4$ binary symbols.}

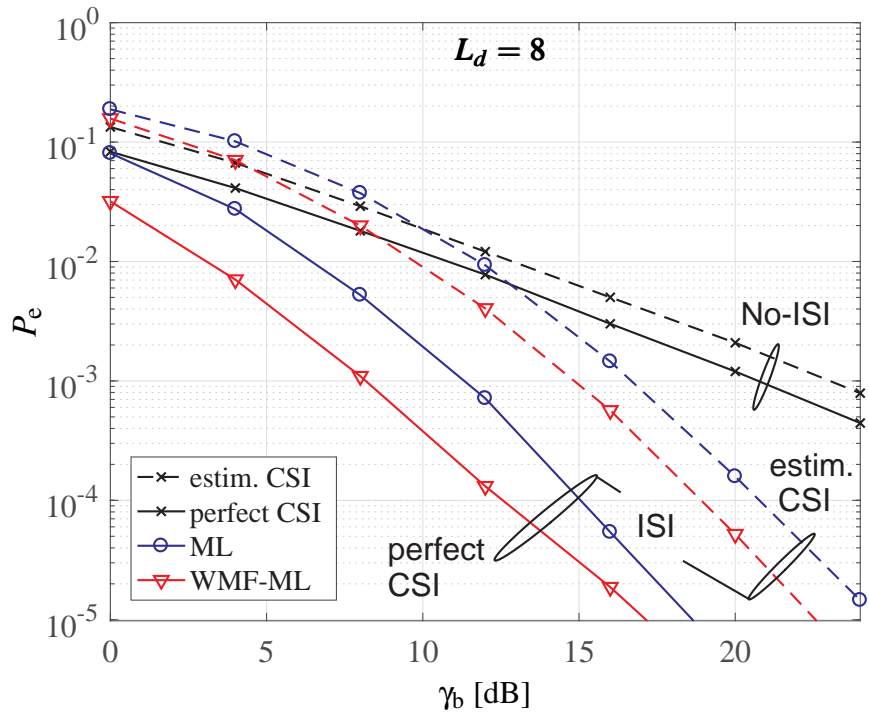
\insfig{0}{scale=1.0}{pict08}{The BER of BPSK vs. SNR over Rayleigh fading
  channel for sequences of $8$ binary symbols.}

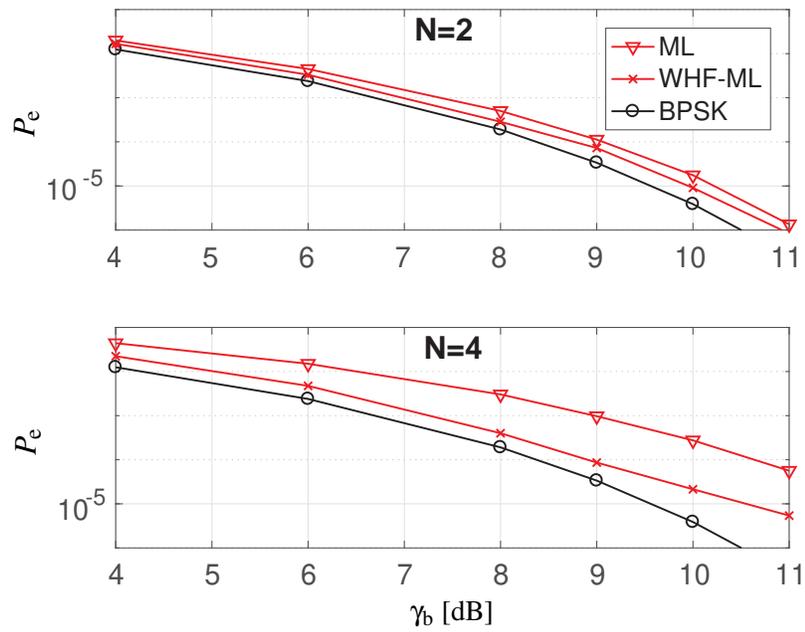
\insfig{0}{scale=0.90}{pict09}{The BER of BPSK vs. SNR over AWGN channel for
  sequences of $2$ and $4$ binary symbols, respectively.}

Lastly, the BER performance of Nyquist modulation and pulse-shape multiplex
modulation transmitting differentially encoded quadrature PSK (QPSK) symbols
over an AWGN channel is compared in \fref{pict10}. It can be observed that
even though the pulse-shape multiplexing suffers asymptotically a 3 dB penalty
in SNR, it reduces the time required for transmitting the whole symbol sequence
to one half.

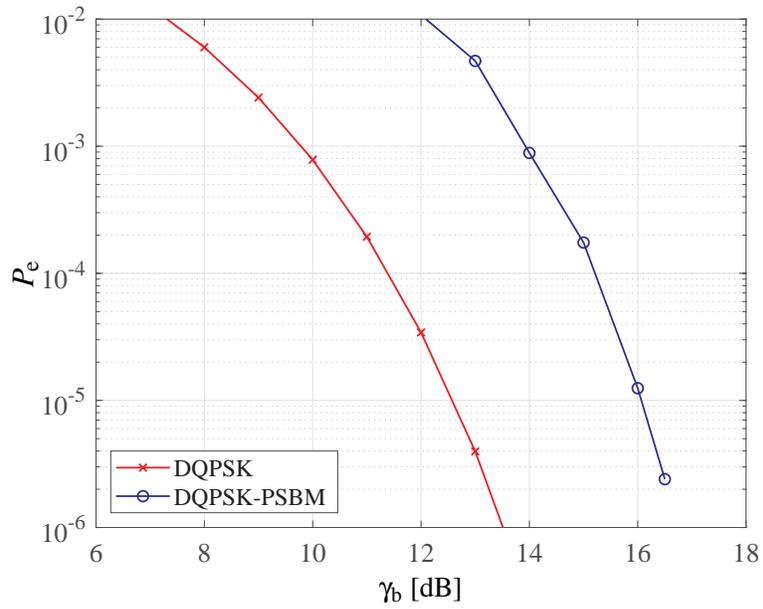
\insfig{0}{scale=0.9}{pict10}{The BER comparison of differentially encoded
  QPSK with Nyquist and pulse-shape binary multiplexing (PSBM) modulation
  transmitted over an AWGN channel.}

\section{Conclusion}

The paper introduced a pulse-shape binary multiplex modulation. Such a
modulation scheme is akin to partial-response signaling, correlative coding,
offset-QPSK modulation and FTN signaling. It combines two data streams under
controlled ISI created by the RRC pulses having $100\%$ roll-off, and
transmitted at twice the Nyquist rate. The ISI analysis showed that this is
unique property among all the roll-off factors being at most $100\%$ and the
packing factors greater than $5\%$. However, the successive samples of additive
noises at the output of matched filter at the receiver become correlated, which
incurs a SNR performance penalty. This penalty could be reduced or even removed
by using more complex sequence-based detection schemes as shown elsewhere in
the literature. The BER performance as well as decoding complexity of the
proposed pulse-shape binary multiplexing modulation scheme is critically
affected by the choice of transmitted sequences. One can consider superposition
modulation with SIC decoding, interleave data symbols with pilot and
zero-symbols to aid channel estimation and data decoding, and also employ
orthogonal spreading sequences to separate the multiplexed data streams. The
numerical results indicate that pulse-shape binary multiplexing can exploit
time-diversity in fading channels to outperform the Nyquist signaling. In
addition, it has been shown numerically that a sequence of differentially
encoded PSK symbols can be transmitted twice as fast by the proposed modulation
scheme compared to canonical Nyquist signaling, although with a 3 dB SNR
penalty over AWGN channels.

\clearpage

\bibliographystyle{ieee}
\bibliography{refer}

\end{document}